# High-resolution Calorimetry in Pulsed Magnetic Fields

Shusaku Imajo[1*], Chao Dong[1,2], Akira Matsuo[1], Koichi Kindo[1], and Yoshimitsu Kohama[1*]

[1] *Institute for Solid State Physics, The University of Tokyo, Kashiwa 277-8581, Japan*

[2] *Wuhan National High Magnetic Field Center and School of Physics, Huazhong University of Science and Technology, Wuhan 430074, China*

We have developed a new calorimeter for measuring thermodynamic properties in pulsed magnetic fields. An instrumental design is described along with the construction details including the sensitivity of a $RuO_2$ thermometer. The operation of the calorimeter is demonstrated by measuring heat capacity of three samples, pure Germanium, $CeCu_2Ge_2$, and κ-$(BEDT-TTF)_2Cu[N(CN)_2]Br$, in pulsed fields up to 43.5 T. We have found that realization of field-stability is a key for measuring high-resolution heat capacity under pulsed fields. We have also tested the performance of the calorimeter by employing two measurement techniques; the quasi-adiabatic and dual-slope methods. We demonstrate that the calorimeter developed in this study is capable of performing high-resolution calorimetry in pulsed magnetic fields, which opens new opportunities for high-field thermodynamic studies.

[*]imajo@issp.u-tokyo.ac.jp

[*]ykohama@issp.u-tokyo.ac.jp

## I. Introduction

The generation of magnetic fields higher than 20 T was first achieved by Pyotr L. Kapitsa approximately a century ago [1]. After Kapitsa's pioneering work, various techniques for producing high magnetic fields have been proposed and established. For instance, the combination of a Bitter electromagnet and a superconducting magnet, known as a hybrid magnet, can generate static magnetic fields up to 45 T for a long time [2]. The electromagnetic flux-compression technique initially invented by S. Chikazumi can produce ultra-high pulsed magnetic fields up to 1200 T, but only for an extremely short time of a few μs [3]. The available field range for condensed matter experiments is constrained by a trade-off between the duration and strength of magnetic fields [4]. Therefore, the experimental investigations in pulsed fields must be performed for a short time; a high-frequency measurement is required. However, performing high-frequency calorimetry has been technically challenging, because the thermometer and the sample need to reach thermal equilibrium within the available time scale.

The first successful heat capacity measurement in pulsed fields was performed by using a long-pulsed magnetic field powered by a flywheel generator [5], demonstrating the validity and the reliability of thermodynamic studies in pulsed fields. Here, the measurement time scale imposed by the pulse duration was 0.1 ~ 1 s with a 60 T long-Pulse Magnet in NHMFL, which enables to apply the quasi-adiabatic method at a measurement frequency of ~50 Hz. In the last decade, the measurement frequency was improved to 1 ~ 2 kHz with the use of a thin-film thermometer [6] and to 1 ~ 10 kHz with applying AC technique [7]. However, the achieved data quality of heat capacity has been strongly limited due to the electromagnetic noise from a rapid field sweep [6-9], and further improvements of the measurement technique have been desired for detailed thermodynamic studies in high magnetic fields.

In this paper, we describe two approaches that can provide accurate and high-resolution heat capacity data in pulsed fields. One is the classical quasi-adiabatic method. Another is the dual-slope method which can collect a large number of data points in a single pulse. We discuss the ideal thermal time constants for both approaches. In the pulsed field experiments, the error originating from the magnetocaloric effect (MCE) and eddy current heating are eliminated by stabilizing the field strength at constant values. We also report temperature and field dependences of resistance of home-made $RuO_2$ thermometers. These developments have led to significant improvements in accuracy and resolution of heat capacity measurements and should serve as a novel guideline for future thermodynamic studies in high magnetic fields.

## II. Apparatus

### A. Design of probe

Figure 1 shows the schematic illustration of our setup which was composed of two parts; the inner probe and the outer double-wall tube. The inner probe was the main part of the setup, which had the calorimeter at the bottom, while the outer double-wall tube worked as a $^3$He refrigerator. The probe and the calorimeter were constructed by non-metallic components to reduce the eddy-current heating and the resultant thermal gradient. For example, the bottom parts of the probe such as a pipe and sealing cap were made by fiber-reinforced plastics (FRP). To maintain a vacuum seal and keep the adiabatic condition, the uncured epoxy for cryogenic use (Nitto Denko, Nitofix SK-229), which can be removed afterwards, was used for the glue between the FRP pipe and the sealing cap. In the middle of the probe, where no high magnetic fields were present, the FRP pipe was connected to the SUS304 pipe with the cured epoxy resin. These vacuum sealings are important for obtaining an adiabatic condition, because the existence of remnants of gas and liquid causes unexpected background contribution and extremely high thermal conductance to the calorimeter. The sealing cap was removed from the inner tube while a sample was mounted on the sample stage. The bottom part of the probe was immersed into liquid $^3$He (Liq. $^3$He) to thermalize the calorimeter at low temperature. Note that Liq. $^3$He played the role of heat bath, which was thermally stable even in high magnetic fields thanks to huge heat capacity of Liq. $^3$He. To provide a stable low-temperature condition for the sample stage (Fig.1), we chose fine manganin wires for the electrical circuits, because their bad thermal conductance can suppress the heat flow from the surroundings to the sample stage. The manganin wires were soldered to 0.1 mm diameter silver wires. The silver wires passed through the FRP wall, which was sealed with the epoxy resin, as displayed in the inset of Fig. 1. One end of the silver wire was thermally coupled to the Liq. $^3$He heat bath and the other end was connected to the Pt-W(8%) relaxation wires. A series of thermal radiation shields made from aluminum-foils were arranged inside the inner probe to thermally separate the sample space from the probe top.

The outer double-wall tube was composed of two tubes. One was the inner FRP tube, and the other was the outer SUS tube, where the vacuum space between two tubes acted as a thermal insulation layer. To accumulate Liq. $^3$He at the bottom of the double-wall, as depicted in Fig. 1, $^3$He gas was introduced from the gas line at the top of the probe. The gas was condensed around the thermal anchor made of SUS 304 by heat exchange with a pumped Liq. $^4$He bath (~1.5 K). The liquefied $^3$He was dropped to the bottom of the double-wall tube and was cooled down to ~500 mK by pumping itself. The outer SUS tube can heat up during field pulse, but the heat cannot transfer to the Liq. $^3$He owing to the vacuum space. This setup can maintain the constant temperature of ~500 mK during the field pulses up to 43.5 T.

*B. Sample platform*

As shown in the photo of Fig. 1, the sample platform was composed of the sample stage, the silver electrode, the heater, the thermometer, and the thermal relaxation wires. The sample was mounted on the sample stage by a small amount of Apiezon N grease. The heat capacity of the sample $C_{sam}$ was obtained by subtracting the background contribution called addenda $C_{ad}$ from the total heat capacity $C_{tot}$. The addenda consisting of the sample platform and the Apiezon N grease was determined in a separate run performed beforehand.

For the sample stage, we selected a single crystal of a TiO$_2$ (rutile) substrate because of the small lattice heat capacity of rutile (Debye temperature ~780 K [10]). The dimension of the substrate was approximately 1*1*0.2 mm$^3$, ensuring easy mounting and de-mounting of samples with a sample weight up to ~10 mg. The film thermometer and heater were directly prepared on the surface of the rutile substrate by baking the commercially available RuO$_2$ pastes, EZ-13 and EZ-14 (PREXNET, TANAKA KIKINZOKU, Japan) at 850°C for 10 minutes. To perform the high-frequency (~ 10 kHz) AC resistance measurements without any phase shift, we adjusted the resistance of the RuO$_2$ thermometer in the range of 100 to 1000 Ω at 1 K by modifying the aspect ratio of the resistor. The temperature and magnetic field dependence of the resistance can be also tuned by the mixing ratio of some different types of RuO$_2$ paste. The calibration of the magnetoresistance was performed under isothermal conditions in Liq. $^3$He or $^4$He. These results as well as the temperature dependence of $C_{ad}$ will be presented in Chapter IV. The silver electrode was composed of the silver paste or the small pieces of silver foil which were electrically connected to the RuO$_2$ thermometer and heaters. The electrodes were soldered to the thermal relaxation wires made of φ50 μm Pt-W8% with eutectic Au-Sn alloy. Each Pt-W wire was connected to the silver wire, which was thermally anchored to the Liq. $^3$He heat bath.

We would emphasize that these selections of materials for the sample stage were based on the concept that the reduction of $C_{ad}$ increases the portion of $C_{sam}$ in $C_{tot}$ and make the S/N ratio better. For example, $TiO_2$ does not contain large nuclear Schottky anomaly under low-temperature and high-field conditions, because it is mainly composed of $^{46}Ti$, $^{48}Ti$, $^{50}Ti$ (87.2% of Ti atom has $I = 0$), $^{16}O$ and $^{18}O$ (99.9% of O atom has $I = 0$). This is an advantage comparing to the sapphire ($Al_2O_3$) substrate includes $^{27}Al$ (100% Al atom has $I = 5/2$). The Pt-W alloy was chosen for the thermal relaxation wire due to its firmness and low thermal conductivity, which suppress the vibration noise and heat leak through the wires.

*C. Electronics*

Since the typical time scale of pulsed magnetic fields is on the order of $10^{0-3}$ ms, which is comparable with the response time of commercial AC resistance bridges, the electronics specific to the high-speed calorimetry is necessary. A block diagram of the electronics used in our experiment is shown in Fig. 2. For quick data collection, PXIe-6124 multifunction data acquisition system (National Instrument) was employed, which was capable of generating excitation current and of detecting the resultant voltage. The sampling rate of 400 kS was used for the data generation and acquisition. The AC voltage $V_{ex}$, whose frequency was 10 kHz, was applied to the thermometer as the excitation current $I_{ex}$, whereas the DC pulsed current was given to the heater for applying heat to the sample platform. To monitor currents applied to them, standard resistors of 0.5 ~ 1 kΩ were put in the electric circuits as shunt resistors. The electrical signals obtained in the measurement were filtered and amplified by using SR560 preamplifiers (Stanford Research). With the measured voltage $V_{th1}$, $V_{th2}$, $V_{h1}$ and $V_{h2}$ in Fig. 2, the resistance of the thermometer and heater were measured by quasi-four-terminal technique $R_{th} = V_{th1}V_{th2}/R_{s1}$ and $R_h = V_{h1}V_{h2}/R_{s2}$, where two wires connecting the electrodes of each resistor bifurcates as four wires at the junctions between the relaxation and Ag wires. The total resistance of the Pt-W relaxation wire and the contact resistance at the junctions, which is typically about 1 Ω, were negligibly smaller than the resistance of $RuO_2$ ~1 kΩ. The thermometer calibration was performed by taking the contribution, and thereby we could neglect the contribution in thermometry. The estimation error of heat was less than 0.2 % by setting high heater resistance over 500 times larger than the resistance of the Pt-W wire.

*D. Magnetic fields*

A non-destructive pulsed magnet can generate high magnetic fields up to ~100 T, but only for a short time of 1–1000 ms. Because of the rapid field-sweep rate ranged from 0.1 to 50 kT/s, the eddy current and/or the MCE induce large undesirable temperature change in pulsed fields. To suppress the temperature instability, we have generated specially designed flat-top magnetic fields by employing the Proportional-Integral-Derivative (PID) feedback technique of the magnetic fields [11]. The home-made PID controller regulates magnetic fields generated by a mini-coil to flatten the time dependence of the magnetic field generated by the main magnet. This technique is capable of stabilizing the field strength up to a sub-second timescale, enabling to measure the relaxation process, and can be used for the studies of heat capacity and Nuclear Magnetic Resonance (NMR) in high magnetic fields.

As the main magnet, we selected a long-pulse magnet installed at International MegaGauss Science Laboratory, Institute for Solid State Physics, the University of Tokyo, which had a large inner bore diameter of 30 mm with the inductance of 60 mH and resistance of 56 mΩ at 77 K. The mini-coil inserted into the main magnet has a 29.3 mm outer bore diameter and 20.0 mm inner diameter with the inductance of 0.9 mH and resistance of 132 mΩ at 77 K. The main magnet was energized by the 51.3 MW DC flywheel generator, and the mini-coil was driven by the four 12-V lead-acid batteries connected in series. The current through the mini-coil was regulated by the IGBT (1MBI900V-120-50, Fuji electric) and the FPGA module (National Instruments USB-7856R).

Figure 3 shows the field profiles of the flat-top pulsed fields. Highly stabilized flat-top pulsed fields up to 43.5 T are generated, where the duration of the flat-top field is 70-500 ms depending on the generated field strength. For example, the 24.3 T pulsed-field has 400 ms flat-top region within ±5 mT, corresponding to a sweep rate of ~0.01 T/s that is several orders of magnitude smaller than those of typical pulsed magnetic fields.

## III. Method

*A. Quasi-adiabatic method*

The quasi-adiabatic method is one of the most useful techniques to measure heat capacity in pulsed magnetic fields owing to the short measurement time-scale [12]. In Fig. 4a, we present typical time dependences of temperature $T(t)$ and power $P(t)$ for a quasi-adiabatic measurement. The requirement of this method is that the time scale of each measurement $t_{\text{measure}}$ is enough shorter than the relaxation time constant $\tau_1$, namely $\tau_1 \gg t_{\text{measure}}$. Under this condition, the heat leak through the relaxation wires is significantly small (not negligible) in $t_{\text{measure}}$, and a simple fitting procedure can be applied for the correction of the heat leak effect. As seen in Fig. 4a, the application of heat $\Delta Q$ induces the temperature increments $\Delta T$ and subsequently leads to the slow thermal relaxation with the time scale of $\tau_1$ to the bath temperature under the quasi-adiabatic condition. Using the Maclaurin series for the exponential function of the Taylor expansion, the slight slope of the relaxation curve can be expressed by the linear time dependence $(-\Delta T/\tau_1)t$ in the case of $\tau_1 \gg t_{\text{measure}}$. As displayed in Fig.4a, the linear extrapolation,

$$T = T_0 + \Delta T - \frac{\Delta T}{\tau_1}\left(t - \frac{\Delta t}{2}\right), \quad (1)$$

, allows us to determine the reasonable $\Delta T$ expected in an adiabatic condition. $\Delta Q$ is precisely estimated by integrating the Joule heating applied to the heater $P$,

$$\Delta Q = \int P dt = \int (V_{\text{h1}} V_{\text{h2}}/R_{\text{s2}}) dt, \quad (2)$$

, where $V_{\text{h1}}$, $V_{\text{h2}}$, and $R_{\text{s2}}$ are the voltage and the shunt resistance shown in Fig. 2. The estimations of $\Delta T$ and $\Delta Q$ permit to determine heat capacity via the following standard formula in the adiabatic condition, $C = \Delta Q/\Delta T$. In our measurement, $\Delta T$ was set in the range of 3-7% of the sample temperature $T$ by regulating $\Delta Q$.

The above-mentioned simple extrapolation is sometimes unusable due to the so-called "$\tau_2$ effect" originating from the poor thermal contact between the sample and the thermometer. Taking the $\tau_2$ effect into account, the time delay of the temperature response must be reconsidered. In our set-up, the sample platform that is thermally well connected to the heater is firstly heated up once the $\Delta Q$ is applied to the heater. After that, the heat is propagated to the sample with a time scale of $\tau_2$. The thermal relaxation between the heater and the sample is also described in a single exponential $\exp(-t/\tau_2)$. In Fig. 4a, we show an example of the time dependence of temperature with the non-negligible $\tau_2$ effect ($\tau_2 \sim 0.1 t_{\text{measure}}$). The behavior can be expressed by the formula convoluted with the $\tau_2$ relaxation,

$$T = T_0 + \Delta T - \frac{\Delta T}{\tau_1}\left(t - \frac{\Delta t}{2}\right) + \Delta T' \exp\left(-\frac{t - \Delta t/2}{\tau_2}\right). \quad (3)$$

As seen in the dotted line in Fig. 4a, the extrapolation of $T(t)$ curve much after $\tau_2$ relaxation can give a reasonable estimate of $\Delta T$. Nevertheless, long $\tau_2$ comparable to $t_{\text{measure}}$ gives an ambiguity to the estimation of $\Delta T$ due to the mixing of the $\tau_2$ effect and the thermal relaxation by $\tau_1$. Therefore, the condition

$\tau_1 \gg t_{measure} \gg \tau_2$ must be fulfilled for the quasi-adiabatic method. In the present measurements, we set $t_{measure}$ about 10 ms sufficiently shorter than $\tau_1 \sim$ 100-1000 ms and longer than $\tau_2 \sim$ 0.1-1 ms.

*B. Dual-slope method*

Performing the quasi-adiabatic method, the condition $\tau_1 \gg t_{measure} \gg \tau_2$ is required as was introduced. Since $\tau_1$ is given by the ratio of $C_{tot}$ to the thermal conductance of the relaxation wire, the requirement $\tau_1 \gg t_{measure}$ is not occasionally fulfilled with the field or temperature changes of the sample heat capacity. In such cases, we can employ the dual-slope method as shown in Fig. 4b [13] that is also called "long-relaxation method" [14]. For this method, the conditions $t_{measure} \gg \tau_2$ and $t_{flat-top} > \tau_1$ are needed, where $t_{flat-top}$ is the time scale of the stabilized field region in pulsed magnetic fields. Using this technique, heat capacity can be measured even in the case $\tau_1 \sim t_{measure} \gg \tau_2$, in which the quasi-adiabatic method cannot be applied. In this method, the stabilization of the base temperature $T_0$ is necessary before the heat application (see Fig. 4b). Once the temperature stabilization is ensured, the constant power is applied to raise the sample temperature. The typical raise temperature of 1.5-3$T_0$ is used in our set-up. Considering the balance of heat quantities at a certain temperature, the increase in temperature is formalized as,

$$P - \int_{T_0}^{T} \kappa(T') dT' = C_{tot} \frac{dT}{dt}\Big|_{up}, \quad (4)$$

where $\kappa$ denotes the thermal conductance of the relaxation wire. After switching off the heater, temperature gradually decrease by the heat release through the relaxation wire. Similar to the heating process, the time dependence of the thermal relaxation curve is expressed with $C_{tot}$ and $\kappa$ as the following formula,

$$-\int_{T_0}^{T} \kappa(T') dT' = C_{tot} \frac{dT}{dt}\Big|_{down}. \quad (5)$$

Hence, heat capacity $C_{tot}$ can be derived from the combination of the two formulae as

$$C_{tot} = P / \left(\frac{dT}{dt}\Big|_{up} - \frac{dT}{dt}\Big|_{down}\right). \quad (6)$$

In practice, the obtained curves are interpolated into arbitral temperatures since the calculation of Eq. 6 requires a data point at the same temperature for both cooling and heating curves. To the best of our knowledge, the correction of $\tau_2$ effect has never been reported for this method, and thus $\tau_2$ should be negligibly small. In the present measurements, we confirm that $\tau_2$ is shorter than or comparable to the sampling rate of thermometry at 10 kHz (i.e., $\tau_2 \leq 0.1$ ms).

## IV. Result

*A. Sensitivity and performance of the RuO$_2$ thermometer*

Figure 5a is a logarithmic plot of the temperature dependence of dimensionless thermometer-sensitivity $|(dR/dT)/(R/T)|$ to compare the homemade EZ-13 and EZ-14 thermometers with Cenox1030 (Lake Shore Cryotronics Inc.) [15] and the commercially available RuO$_2$ resistor chips, RK73B-1H-103 and RK73B-1E-202 (KOA) [16], which have different $R$ values. To take account of individual differences in thermometers, the data are illustrated as band-like curves estimated by the superposition of multiple curves. From our experiences, the sensitivity more than 0.2 is needed for the heat capacity measurements in pulsed fields, as denoted by the black dashed line. The EZ-13 and EZ-14 resistors at low temperatures meet the required thermometer-sensitivity as well as other thermometers.

In Fig. 5b, we present the magnetoresistance of the EZ-14 resistor. The magnetoresistance of the EZ-13 (not shown) is similar to that of the EZ-14 resistor. As seen in Fig. 5b, the EZ-14 resistor shows relatively small and monotonic field dependence, which is a big advantage for high-field calorimetry, while

Cernox thermometers unfortunately exhibit non-monotonic and large magnetoresistance [15]. The monotonic magnetoresistance can be easily corrected. The normalized magnetoresistance, $\{R(H)-R(0\ T)\}/R(0\ T)$, for the EZ-14 is 5% at 1.4 K and 20 T, leading to 2-4% errors in $\Delta T/T$, unless the magnetoresistance calibration is done. This error is smaller than errors reported in Cernox thermometers ($\Delta T/T \sim 5$ % at 2 K and 20 T) [15].

*B. Examples of heat capacity data measured by the constructed calorimeter*

Figures 6a-c and 6d-f display representative examples taken with the quasi-adiabatic and dual-slope methods, respectively, where temperature (blue), magnetic field (red), and applied heater power (green) are plotted as a function of time. The total heat capacity $C_{tot}$ is determined by the analyses described in Chapter III by using the time dependence of temperature and power. The sample heat capacity $C_{sam}$ (hereinafter called $C_p$) is obtained by subtracting addenda contribution $C_{ad}$. Typical addenda heat capacity in our calorimeter is also measured by the quasi-adiabatic method which is plotted as $C_{ad}/T$ vs. $T^2$ in Fig. 6g. Below 3 K, the small magnetic field dependence of $C_{ad}$ is observed. This is attributable to magnetic impurities lurked in the addenda. The field dependence is roughly described by the Schottky-type anomaly including small effective internal fields. The anomaly in $C_{ad}$ is visible only below 5 T because it is strongly broadened and shifts to a high-temperature above 5 T. The zero field values of $C_{ad}$ are ~15 nJ/K at 1 K and ~500 nJ/K at 5 K, and the values above 10 T are ~8 nJ/K at 1 K and ~500 nJ/K at 5 K.

To evaluate the reliability of the calorimeter, we firstly present the heat capacity data of 22.6 mg polycrystalline germanium at 0 T (blue) and 43.5 T (red) measured by the quasi-adiabatic method in Fig. 7a [17]. The present result is compared with the reported data (the black dotted line) [18]. No obvious field dependence between the 0 T and 43.5 T data is observed, which is reasonable because a natural Ge does not possess electronic magnetic moments. The nuclear Schottky anomaly is also expected to be small due to the great abundance of Ge nucleus with $I = 0$ (natural abundance 92.3 %). We confirm that the deviation of our data from the reference, namely the error, is within 5-10 %. These facts indicate that our calorimeter is capable of measuring the absolute values of $C_p$ in pulsed magnetic fields. Figure 7b is the data of the heavy-fermion antiferromagnet $CeCu_2Ge_2$, partly derived from the results shown in Figs. 6a-c. Here, the $CeCu_2Ge_2$ single crystal of 138.0 µg was used. Comparing the present zero-field data (red) with the data reported in Ref. [19] (black), we again confirm that our calorimeter gives reliable and high-resolution heat capacity data by using the high-speed quasi-adiabatic method. The high resolution permits to quantitatively discuss thermodynamic properties in high fields.

In Fig. 7c, we show the $C_p T^{-1}$ vs. $T^2$ derived from the $T(t)$ data in Figs. 6d-f. The sample measured herein is the quasi-two-dimensional organic superconductor $\kappa$-$(BEDT-TTF)_2Cu[N(CN)_2]Br$. The weight of a single crystal of this salt was 410.5 µg. For comparison, we also show the previous $C_p T^{-1}$ data of the same single crystal measured by the conventional relaxation method [20]. Since $C_{tot}$ is relatively small ($C_{tot} \sim 10$ nJK$^{-1}$ at 1 K) and lead to shorter $\tau_1 \sim 30$ ms below 2 K, we employed the dual-slope method instead of the quasi-adiabatic method. By the quasi-adiabatic method (the inset of Fig. 6e), we confirmed that $\tau_2$ below 2 K is enough shorter than 0.1 ms. At 0 T, the present data (the black boxes) reproduce well the reported data (the red line). This demonstrates that the dual-slope method is also operative with the present calorimeter. With increasing field intensity up to 36.1 T, heat capacity increases due to the suppression of the superconductivity. The subtle discrepancy in the normal state $C_p$ data measured in $H\|b$ (the blue line [20]) indicates that the upper critical field of the sample is higher than 36.1 T for the present field orientation ($H\|a$), which is consistent with the persistence of the superconductivity up to 40 T [20].

## V. Discussion

In this work, we have developed a calorimeter for high-resolution calorimetry in pulsed magnetic fields. As seen in Fig. 5a, it is found that the sensitivities of Cernox1030 and RK73B-1H-103 are better than those of EZ-13 and EZ-14. However, the former bare chip thermometers are not suitable for pulsed field experiments because they contain $Al_2O_3$ in their substrate and the absolute values of resistance are rather large for high-frequency measurements. Therefore, the EZ-13 and EZ-14 resistors are used for our calorimeter. As displayed in Fig. 6g, we find that $C_{ad}$ of our calorimeter is 3-4 times smaller than that of commercially available calorimeters (~200 nJ/K at 2 K) [21] and is almost comparable to home-made high-resolution relaxation calorimeter used in steady magnetic fields [22]. The error is within 10% below 3 K, as shown in Figs. 7a, 7b, and 7c. The resolution of our method depends on the applied excitation voltage, the temperature increment $\Delta T$, the resolution of the electronics, and the electric noise. Nonetheless, the typical resolution of the constructed calorimeter can reach 1 nJ/K at 1 K, which is better than that of commercially available calorimeters (~2 nJ/K at 2 K) [21] as well as that of previous pulsed-field calorimetry [6-9]. This resolution is achieved owing to the highly stabilized flat-top field, the small $C_{ad}$, and the sensitive $RuO_2$ thermometer. We find that even a tiny fluctuation of field strength can induce large temperature instability, which critically disturbs high-resolution heat capacity measurements. For example, in Fig. 6e, the temperature change in the field-changing part other than the flat-top region is quite large for the evaluation of $\Delta T$ and $dT/dt$.

Additionally, it should be noticed that the constructed calorimeter can measure not only the heat capacity in the flat-top field region but also the MCE on the way changing the field strength. The MCE is the spontaneous temperature change by applying magnetic fields that are caused by the entropy change as a function of magnetic field. This technique is sensitive to the magnetic-field-induced phase transitions, since a phase transition is accompanied by entropy change. In some specific cases, where the temperature change is drastic like $1^{st}$-order transitions, entropy change concomitant with the transition can be precisely evaluated, since the rapid temperature change in the short time window can be regarded as an adiabatic process. In our recent work for the metamagnetic $1^{st}$-order transition of $UTe_2$ investigated by the present calorimeter [23], we succeed in obtaining the absolute value of entropy change accompanied by the transition from the MCE because the transition is completed only in 10 ms in pulsed fields. Thus, it should be emphasized that the present calorimetry is useful not only for the heat capacity measurements but also for the MCE measurements.

Finally, we note the time efficiency of the present measurement techniques. For earlier pulsed-field calorimetry [6-9], the data points of heat capacity are sometimes insufficient to discuss physical properties in detail, because only few data points were obtained in each field pulse. However, owing to small $\tau_2$, our calorimeter can collect 10-50 data points for each field pulse by employing the quasi-adiabatic or the dual-slope method. This makes the experimental investigation in pulsed-field easy and enables to complete a data set of heat capacity as functions of temperature and magnetic fields. In the present research, we generate 40 T pulsed fields every few hours; therefore, it is possible to obtain a full data set of $C_p(T, H)$ up to 40 T in the temperature range between 0.5 and 5 K only in several days. The present technique can provide enough data points for detailed discussions of physical properties in a reliable timeframe for condensed matter research.

## VI. Conclusion

In this article, we report the development of the calorimeter customized for heat capacity measurement in pulsed magnetic fields. Constructing the probe and calorimeter with non-magnetic and non-metallic components as much as possible suppresses the large temperature change coming from the magnetocaloric effect and the eddy current. The highly stabilized flat-top fields generated by the combination of the mini-coil and main magnet provide good temperature stability sufficient for heat capacity measurements. By employing the quasi-adiabatic and dual-slope methods, we have succeeded in establishing high-speed and high-resolution calorimetry in pulsed magnetic fields. Comparing the present results with earlier works, we confirmed that the developed calorimeter allows a quantitative discussion of high-field thermodynamic properties with unprecedented precision.

**Acknowledgements** We thank Dr. T. Ebihara (Shizuoka University) for providing a single crystal of $CeCu_2Ge_2$ measured in this study and Dr. K. Matsui (ISSP, the University of Tokyo) for constructing the mini-coil to generate the flat-top fields.

**Figure 1** Schematic illustrations of the probe and calorimeter.

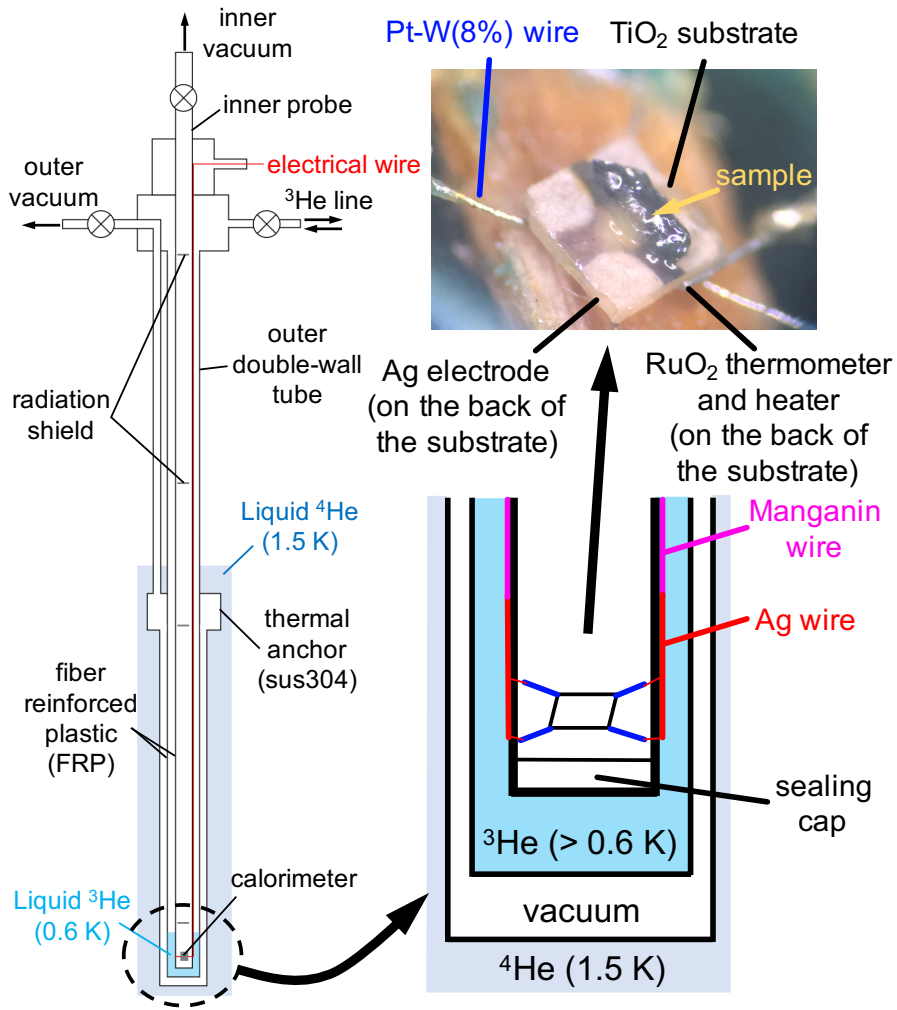

**Figure 2** A block diagram of the electronics used for the high-speed calorimetry in pulsed fields. The gray block belongs to the mini-coil installed inside the main pulse magnet.

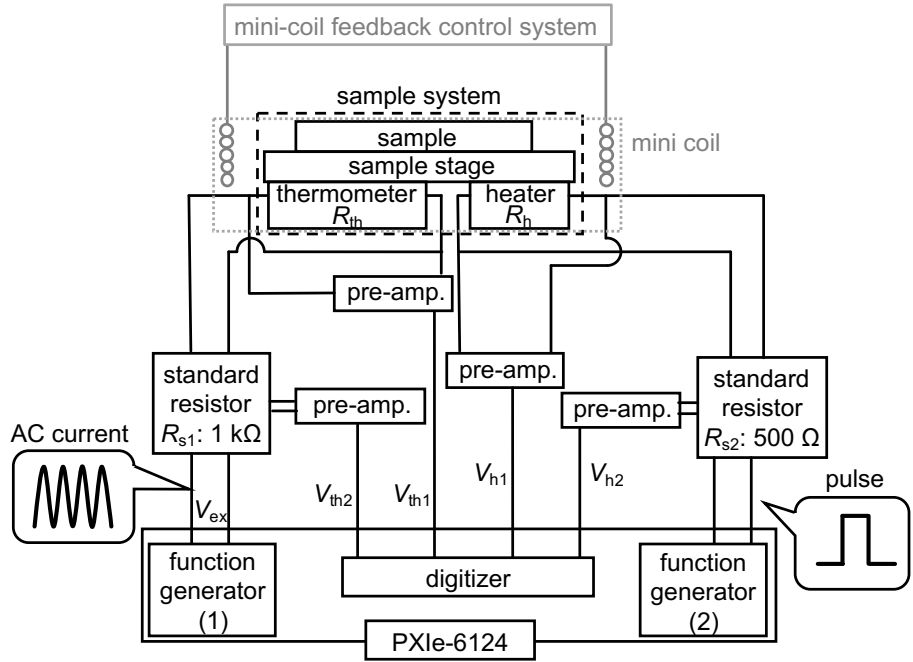

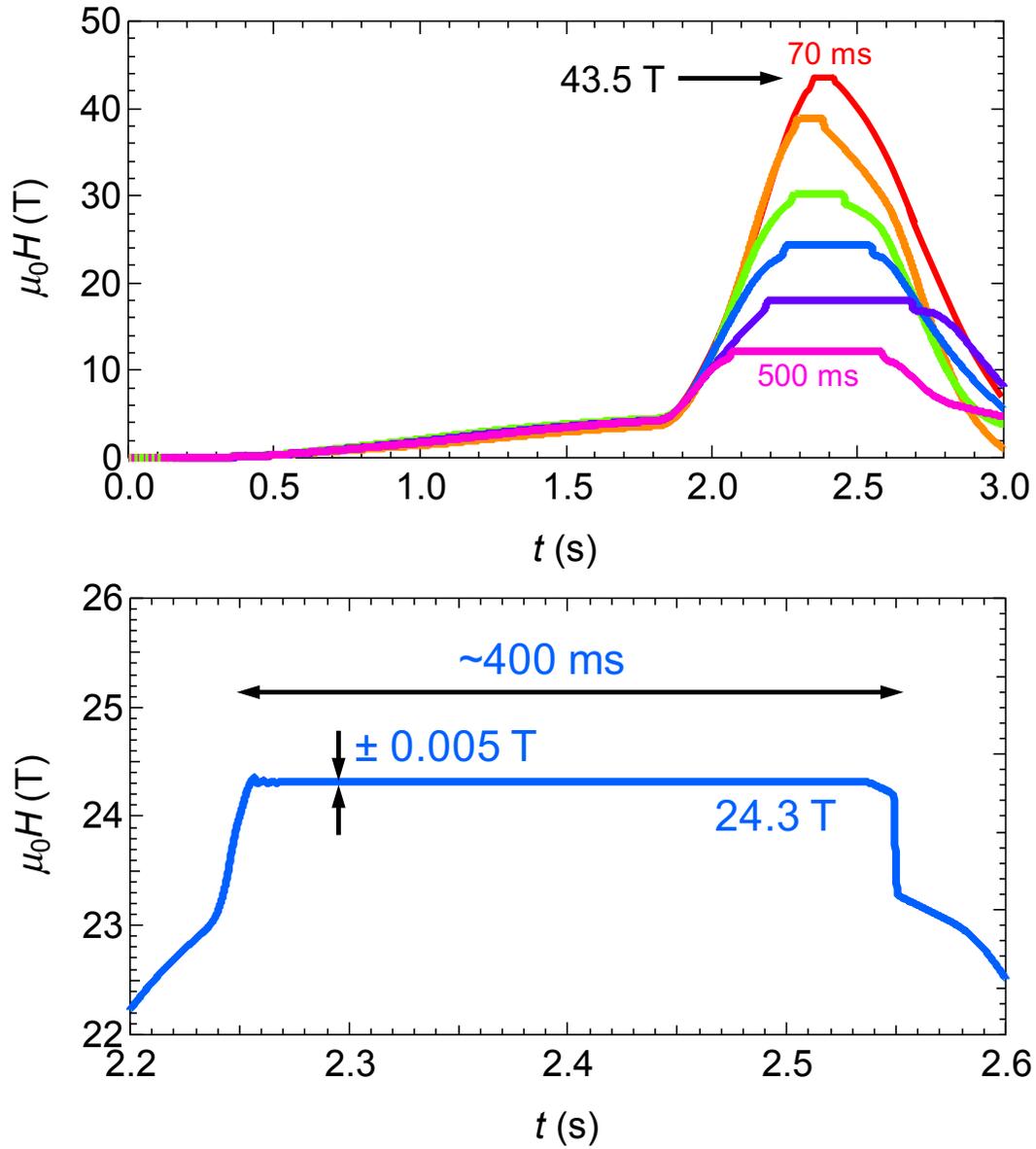

**Figure 3 (a)** Field profiles of pulsed magnetic fields generated by the constructed flat-top system. **(b)** The expanded graph of the 24.3 T shot around the flat-top region.

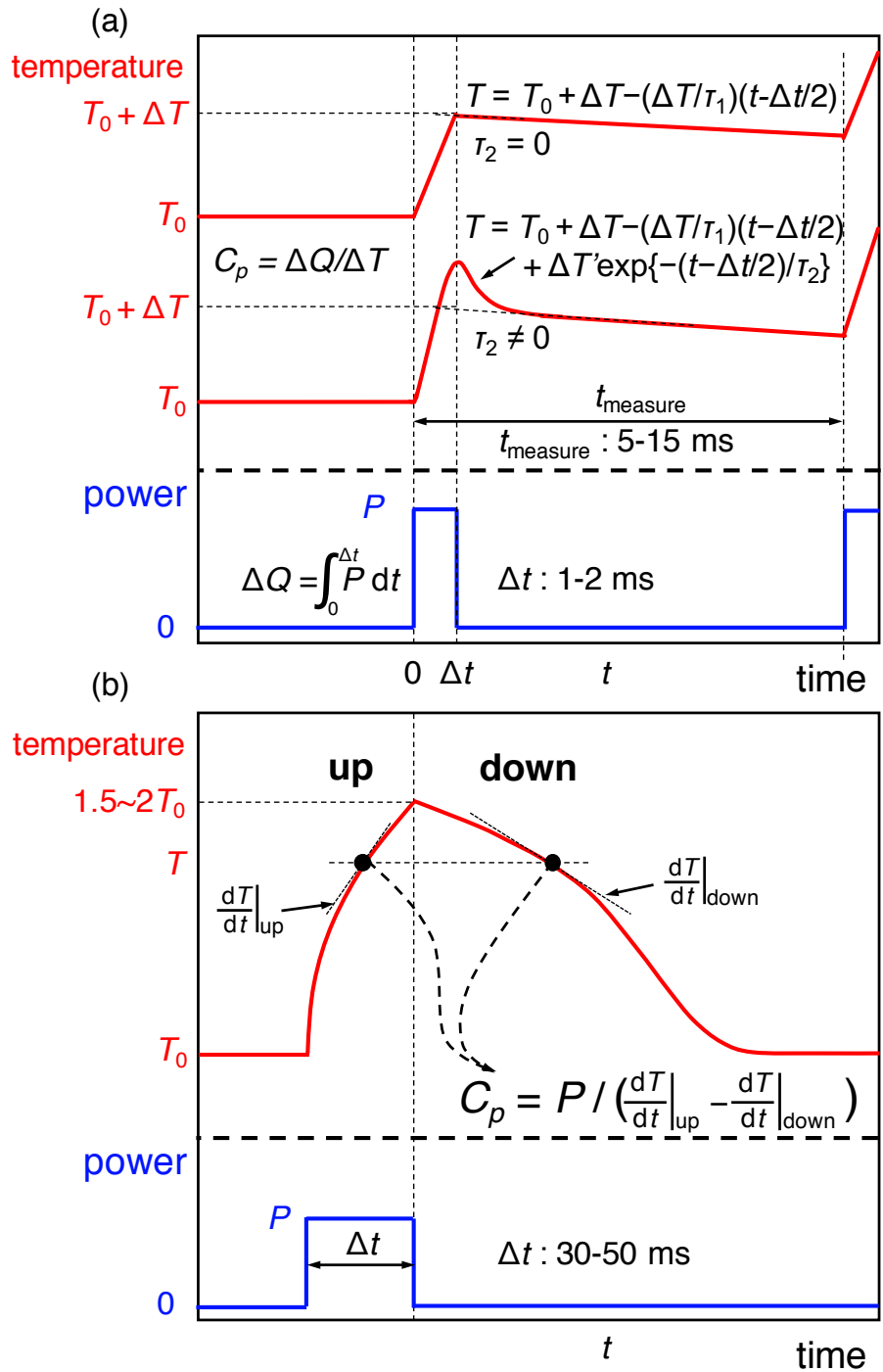

**Figure 4 (a),(b)** Schemes of time dependence of temperature $T$ and applied power $P$ of **(a)** the quasi-adiabatic method and **(b)** the dual-slope method, respectively. The upper red curve in (a) is the case where $\tau_2=0$ whereas the middle red curve is for $\tau_2\neq 0$. The bottom blue curve indicates the time-power profile.

**Figure 5 (a)** Dimensionless sensitivity $|(dR/dT)/(R/T)|$ of some thermometers as a function of temperature. The black dashed line is a rough standard of the sensitivity required for the present calorimetry. **(b)** Reduced magnetoresistance of the EZ-14 thermometer at various temperatures. For clarity, we add vertical offsets signified by the dotted lines to the curves.

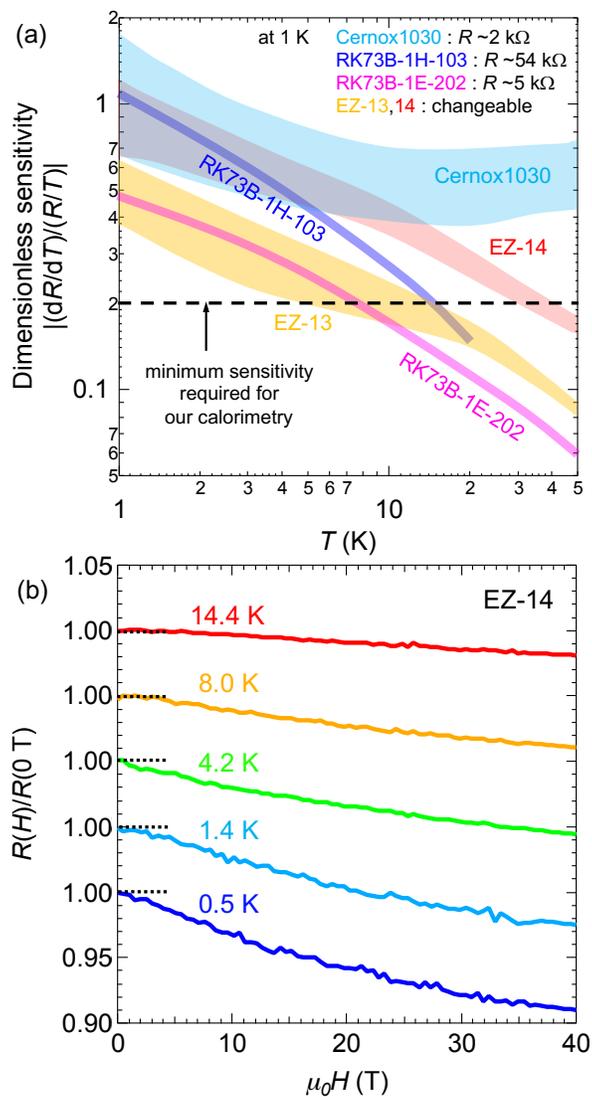

**Figure 6 (a)** Time dependence of temperature and field for the typical case of the quasi-adiabatic method. **(b)** The expansion of (a) around 2.3 s. **(c)** The further expansion of (b) and applied power around 2.29 s. The red dotted curves are fits to the formula introduced in the main text. **(d),(e),(f)** Practical time-field(d)/temperature(e)/power(f) profiles of the dual-slope method. The temperature $T_0$ indicates the base temperature. The insets of (e) and (f) are the enlargement of the quasi-adiabatic method done around 2.3 s for comparison. **(g)** Typical addenda heat capacity of 0 T and 17.5 T as a function of temperature.

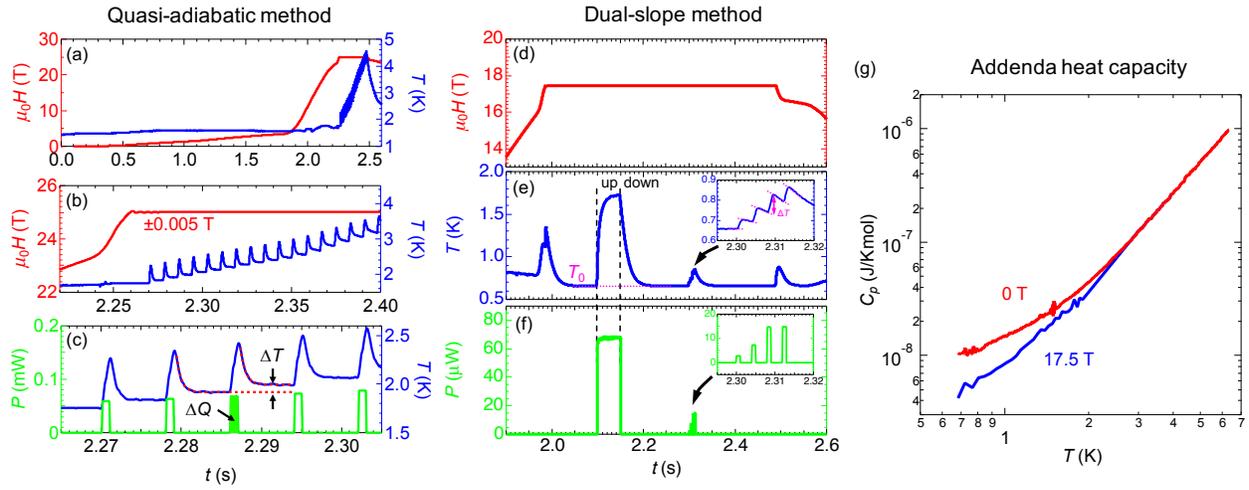

**Figure 7 (a)** Heat capacity of pure Germanium at 0 T (blue) and 43.5 T (red) measured by the present calorimeter. The black dotted line represents the reference data reported in Ref. [17]. **(b)** Temperature dependence of heat capacity of $CeCu_2Ge_2$, partly derived from the profile shown in Figs. 6a-c. The data reported in Ref. [19] (black) are well consistent with the present data at 0 T (red). **(c)** Low-temperature heat capacity of the organic superconductor κ-(BEDT-TTF)$_2$Cu[N(CN)$_2$]Br up to 36.1 T as the $C_pT^{-1}$ vs. $T^2$ plot. The red and blue curves denote the heat capacity data of 0 T and 7 T of a field perpendicular to the conducting plane (normal state) reported in Ref. [20].

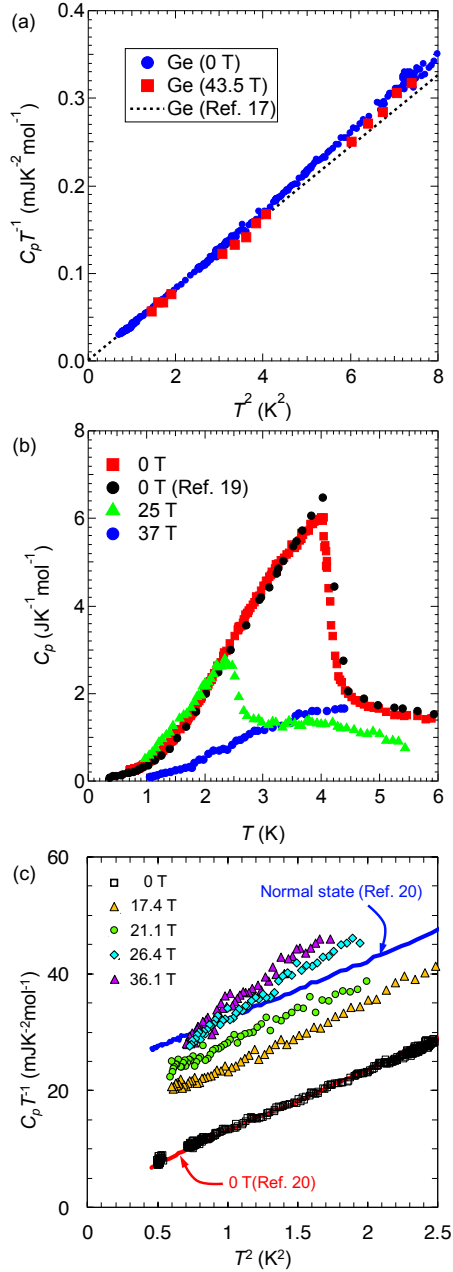